\documentclass{article}

\usepackage[eandd, final]{neurips_2026}

\usepackage[utf8]{inputenc} 
\usepackage[T1]{fontenc}    
\usepackage{hyperref}       
\usepackage{url}            
\usepackage{booktabs}       
\usepackage{amsfonts}       
\usepackage{nicefrac}       
\usepackage{microtype}      
\usepackage{xcolor}         
\usepackage{amssymb}
\usepackage{enumitem}
\usepackage{multirow}
\usepackage{algorithm}
\usepackage{algpseudocode}
\usepackage[table]{xcolor}   
\usepackage{graphicx}
\usepackage{tcolorbox}
\usepackage{subcaption}

\makeatletter
\renewcommand{\@noticestring}{}
\makeatother

\title{MELT: A Behavioral Trace Dataset for High-Risk Memecoin Launch Detection}

\author{%
  Sihao Hu, Selim Furkan Tekin, Yichang Xu, Ling Liu \\
  School of Computer Science\\
  Georgia Institute of Technology, Atlanta, USA  \\
  \small \texttt{\{shu335,stekin6,xuyichang,ll72\}@gatech.edu}
}

\begin{document}

\maketitle

\begin{abstract}

Launchpads have become the dominant mechanism for issuing memecoins, exposing investors to a new class of high-risk launches that existing rug-pull detection methods cannot capture. We argue that detecting these threats requires structured behavioral traces that underlie raw heterogeneous blockchain data, \emph{i.e.}, how insiders accumulate, coordinate, and unwind positions. To enable such analysis, we introduce MELT (\textbf{ME}mecoin \textbf{L}aunch \textbf{T}race), the first behavioral trace dataset for analyzing and detecting high-risk memecoin launches on Solana. MELT covers 41k+ memecoin launches with 200M+ transactions parsed into typed behavioral records that distinguish swaps, wash trades, transfers, and mints. Beyond per-account behaviors, MELT contributes bundle-trace data that links accounts controlled by the same entity, revealing that, on average, 36.5\% of token supply is held by coordinated accounts, a concealment strategy that disguises the true ownership concentration from unsuspecting buyers. On top of these traces, MELT provides 122 behavioral features and risk-level annotations, enabling supervised learning at a population scale. We benchmark representative ML models on the high-risk launch detection task. Integrating their predictions into a simple memecoin selection strategy reduces investment loss significantly, demonstrating that behavioral traces can be translated into risk mitigation. Our dataset and code is available at \url{https://github.com/git-disl/MELT}.

\end{abstract}

\section{Introduction}

On January 17, 2025, U.S. President Donald Trump launched the \$TRUMP token on the Solana blockchain~\citep{trump_wikipedia}. Within 60 hours its market capitalization surged to roughly \$15 billion, briefly congesting the Solana network and driving daily DEX trading volume to a record \$39.2 billion~\citep{helius_trump_weekend}. The event brought a niche phenomenon into public view: memecoins, digital tokens that embody Internet memes with no intrinsic utility, have nonetheless become a focus of speculative trading at a scale capable of reshaping on-chain market dynamics.

The infrastructure behind this growth is the \emph{launchpad}, a no-code interface that lets anyone create a memecoin within minutes. By October 2025, the largest launchpad, Pump.fun, had issued 12.8 million memecoins on Solana, accounting for roughly half of all tokens ever created on the chain~\citep{galaxy2025pumpfun}. When created, a memecoin is automatically initialized through a launchpad sale at deterministically increasing prices. Once the sale reaches a fixed threshold, the accumulated liquidity is migrated to a Decentralized Exchange (DEX) for public trading. This automated pipeline removes the technical barriers to issuing a memecoin, including those that previously deterred fraudulent operators at scale~\citep{coindesk2025pumpfunfraud}.

\textbf{Related Work.} Launchpads give rise to a new class of threats that prior detection methods are not designed to capture. Existing methods target DEX-based rug pulls~\citep{cernera2023token,mazorra2022not,yaremus2025detecting}, where creators directly own the liquidity pool and can withdraw its assets at will. They therefore rely on pool-level operational signals, such as token burns and sudden liquidity withdrawals. This assumption breaks in the launchpad setting: the liquidity pool is held entirely by the launchpad protocol, leaving creators with no direct control over it but with substantial first-mover and coordination advantages during the sale. As illustrated in Figure~\ref{fig:toy_example}, the threat instead originates during the launchpad sale itself, where insiders (developers and affiliated accounts) accumulate tokens at the lowest price tiers and unwind them after migration, draining base assets and collapsing prices for later buyers. A few recent studies have begun to examine launchpad markets~\citep{voinea2025pump,li2025trust}, but they offer only economic or sociological analyses without systematically detecting this new class of threats.

The core signals for detecting these high-risk memecoins therefore lie in launchpad-stage behavioral traces. However, such traces cannot be directly read off raw on-chain data, for three reasons. First, raw transactions are heterogeneous low-level program invocations whose formats vary across interacting programs and carry no explicit semantics of higher-level user behaviors. Second, behavioral signals such as holding concentration emerge only after aggregating and engineering features over these traces. Most importantly, insiders actively obscure their coordination by operating through multiple accounts that appear to act independently, deliberately concealing cross-account coordination from on-chain analysis.

\textbf{Scopes and Contributions.} To open the new problem for systematic study, we introduce MELT (\textbf{ME}mecoin \textbf{L}aunch \textbf{T}race), the first behavioral trace dataset for analyzing and detecting high-risk memecoin launches on Solana. MELT is constructed over a period centered on the \$TRUMP launch (Dec. 1, 2024 – Mar. 1, 2025), capturing the full memecoin market speculative cycle of growth, peak, and cooling within a single coherent period. It covers all 41,470 memecoin launches that complete a four-stage lifecycle (creation → launchpad sale → migration → DEX trading), with over 200 million on-chain transactions extracted into high-level behavioral records.

Beyond per-account behaviors, MELT contributes bundle-trace data collected from various sources including off-chain data. These bundle traces effectively link accounts likely controlled by the same entity, revealing that 36.5\% of token supply that appears to be held by independent accounts is in fact controlled by coordinated accounts with concentrated ownership. This reflects a deliberate concealment strategy that disguises true ownership concentration from unsuspecting buyers and is invisible in raw transactions. 

On top of these behavioral traces, MELT provides 122 behavioral features across five groups that characterize different aspects of a memecoin launch, paired with a configurable risk-level annotation method that labels each memecoin launch based on its post-migration market trajectory. Together, MELT enables ML models to learn the mapping from launchpad-sale behavior traces to downstream risk outcomes, supporting proactive detection and alerting of high-risk memecoins before their DEX trading begins.

To test the practical utility of MELT, we benchmark a suite of representative ML models on the high-risk memecoin detection task and investigate whether the dataset can be translated into actionable risk mitigation signals. Experiments show that integrating the risk prediction results into trading decisions can reduce investment losses by up to 34 percentage points, confirming its practical defensive value.

\vspace{-0.25cm}
\section{Background}
\vspace{-0.1cm}
\label{sec:background}


\textbf{Account and Transaction.} Solana stores all on-chain state in \emph{accounts}, each identified by a unique address~\citep{solana_docs_accounts}. Users own accounts that hold balances of \emph{SOL} (Solana's native cryptocurrency) and other tokens. A \emph{transaction} executes one or more operations automatically, \textit{i.e.}, either all operations succeed or none take effect~\citep{solana_docs_transactions}. 


\textbf{Memecoin and Mint Address.} Each memecoin on Solana is uniquely identified by a \emph{mint address}, which we use throughout this paper as the global identifier of a memecoin. Throughout this paper, we use \emph{memecoin} to refer to a specific token type, and \emph{token(s)} to refer to the actual balances or units held by users.

\begin{figure}[tbp]
\centering
\begin{minipage}[b]{0.45\textwidth}
    \centering
    \includegraphics[width=\textwidth]{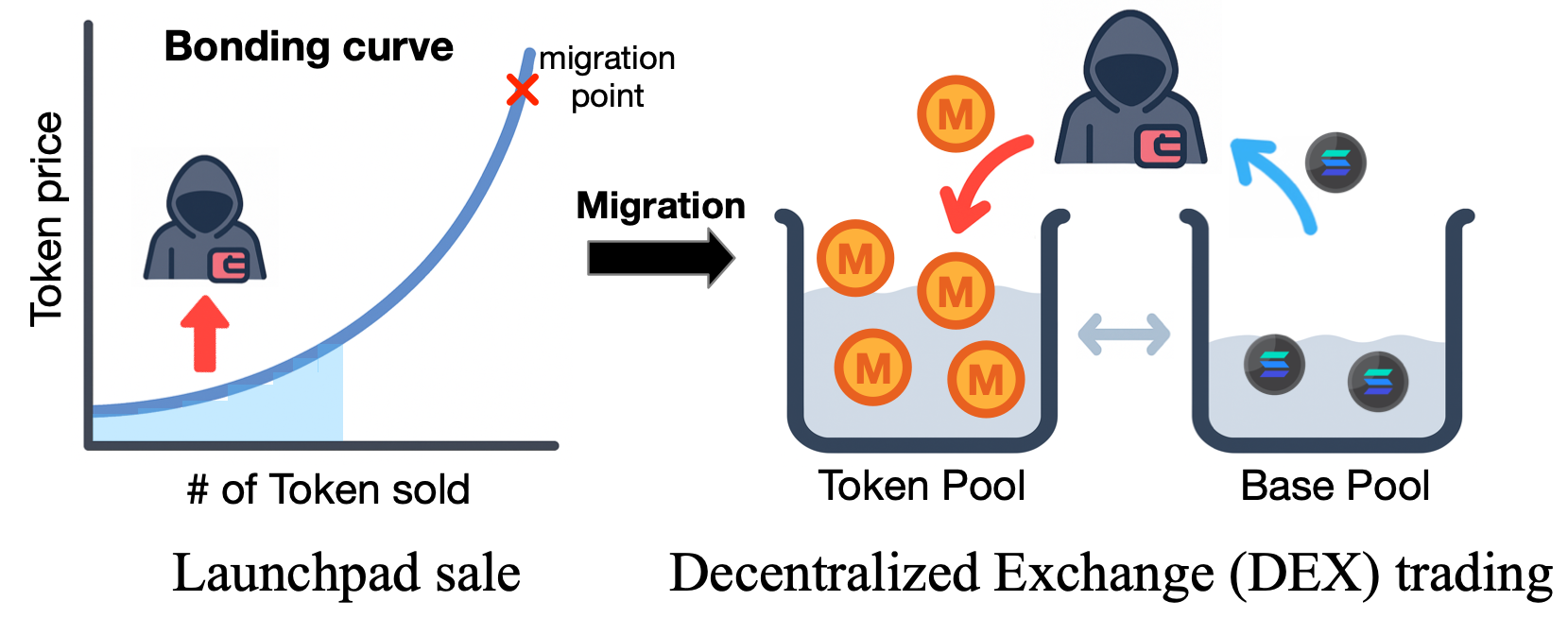}
    \subcaption{Insider accumulation and unwind.}
    \label{fig:toy_example}
\end{minipage}
\hfill
\begin{minipage}[b]{0.54\textwidth}
    \centering
    \includegraphics[width=\textwidth]{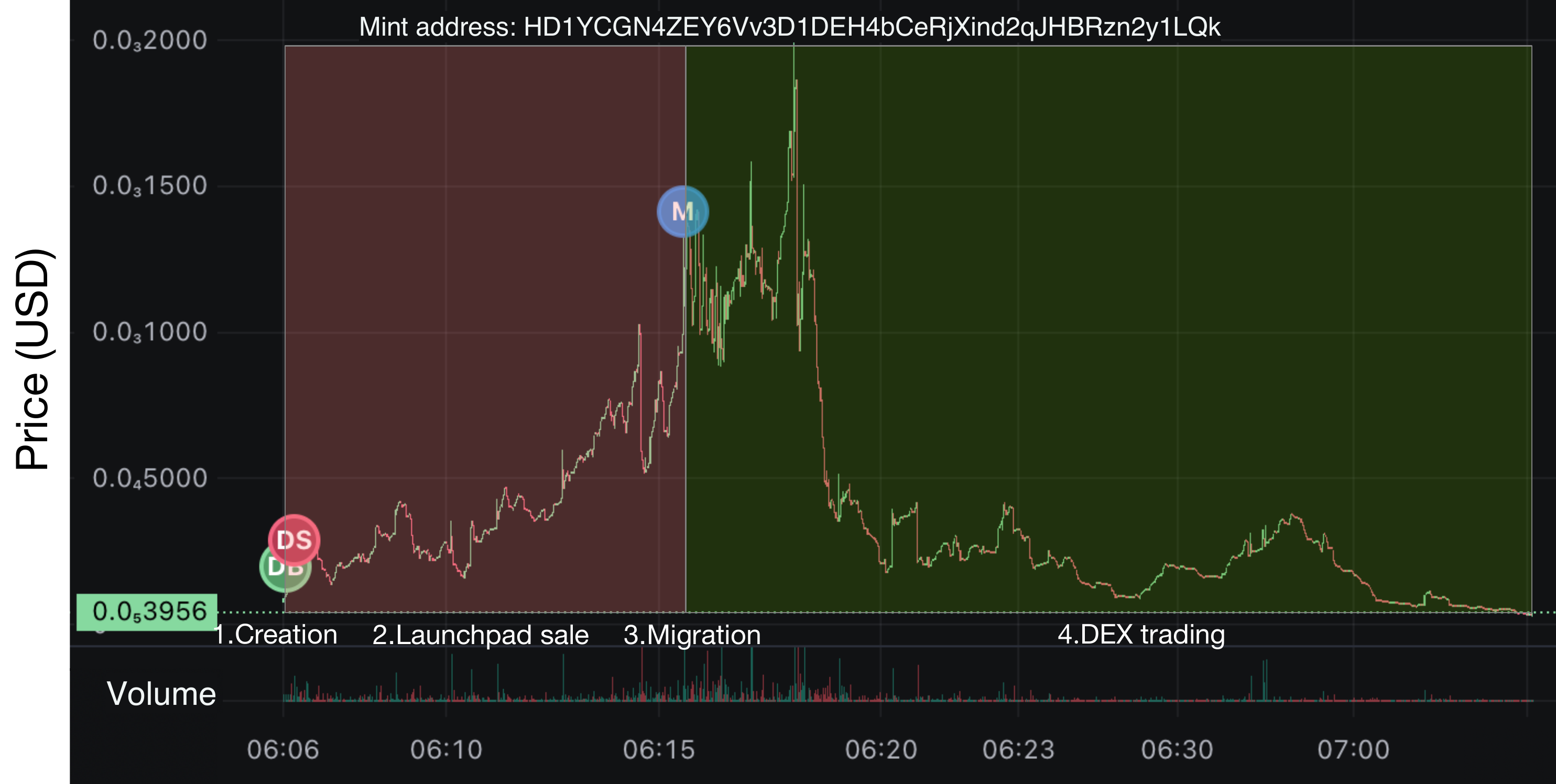}
    \subcaption{A real high-risk memecoin trajectory.}
    \label{fig:case_study}
\end{minipage}
\caption{\textbf{Lifecycle of a memecoin.} 
\textbf{(a)} Insiders accumulate substantial token holdings during the launchpad sale and unwind them during DEX trading, draining base assets and causing price declines. 
\textbf{(b)} A market trajectory of a real memcoin: the developer pre-purchases 97.5M tokens at the earliest bonding-curve tier (\textbf{Creation}). Prices rise as later buyers enter (\textbf{Launchpad sale}). At 80\% sold, the launchpad migrates remaining memecoin tokens and collected base tokens (SOL) to a DEX liquidity pool (\textbf{Migration}), opening public trading with broader exposure (\textbf{DEX trading}). As liquidity deepens, insiders unwind their holdings and collapse the price.}
\vspace{-0.4cm}
\end{figure}

\textbf{Bonding Curve.} During the launchpad sale, users buy or sell tokens directly with the launchpad. The price movement follows a \emph{bonding curve} algorithm: the price depends only on how many tokens have already been sold, rising in fixed steps as cumulative sales increase~\citep{dydx2024bondingcurve}. As shown in Figure~\ref{fig:toy_example}, the first buyers therefore pay the lowest prices, and each subsequent tier is more expensive, an asymmetry that gives insiders a substantial cost advantage.

\textbf{Developers and Insiders.} The \emph{developer} of a memecoin is the account that creates it on the launchpad. We use \emph{insiders} more broadly to refer to the developer together with any coordinated accounts that share informational or operational advantages. Such participants exploit the bonding curve's first-mover advantage to acquire tokens at the lowest price tiers.

\textbf{Decentralized Exchange (DEX) and Liquidity Pool.} After the launchpad sale completes, the memecoin moves to a Decentralized Exchange (DEX) for public trading with broader market exposure. DEXs trade tokens through liquidity pools~\citep{uniswap2018v1overview} that hold paired reserves (e.g., tokens and SOL), where the price is determined by the reserve ratio, \textit{e.g.}, if the pool holds $x$ tokens and $y$ SOL, the token price is $p=y/x$. As a result, selling a large amount of tokens at once can cause the price to fall sharply.

\textbf{The Lifetime of a Memecoin Launch:} Figure~\ref{fig:case_study} illustrates the typical lifetime of a high-risk memecoin with the creation, launchpad sale, migration and DEX trading stages.  Among all the memecoins, less than 2\% ever reach the migration stage~\citep{smithii_pumpfun_2025}. Memecoins that fail to migrate remain confined to the launchpad with negligible trading activity, posing limited risk to the broader investor base. We therefore focus exclusively on memecoins that successfully migrate to a DEX, since they gain sufficient market exposure, making high-risk ones the primary source of financial losses.

\section{Dataset Construction}
\label{sec:dataset}

MELT tracks memecoins issued through Pump.fun from Dec 1, 2024 to Mar 1, 2025, centered on the high-profile launch of the \$TRUMP token on Jan 17, 2025. This event-anchored design captures three market regimes within a single coherent window:
\textbf{(i) Growth phase (Dec 2024 – mid Jan 2025)}: Pump.fun's daily token creation accelerated amid post-election crypto optimism. 
\textbf{(ii) Peak speculation phase (mid Jan – early Feb 2025)}: anchored by the consecutive launches of \$TRUMP (Jan 17) and \$MELANIA (Jan 19) alongside the U.S. presidential inauguration (Jan 20), producing the highest-velocity launchpad trading activity to date.
\textbf{(iii) Cooling phase (early Feb – Mar 2025)}: speculative attention gradually diffused as early insiders unwound positions. The cooling was sharply accelerated by the \$LIBRA collapse on Feb 14, 2025, which extracted \$4.5B market cap within hours, eroding public trust and draining liquidity from the memecoin market.

The construction pipeline consists of five stages: (1) memecoin launch collection, (2) transaction-to-behavior extraction, (3) bundle trace identification, (4) feature engineering, and (5) risk-level annotation. We elaborate each below.


\begin{table*}[tbp]
\centering
\caption{Data structures for example entries of a memecoin launch, a transaction after extraction, a bundle trace, and an annotation label.}
\vspace{-0.2cm}
\includegraphics[width=14cm]{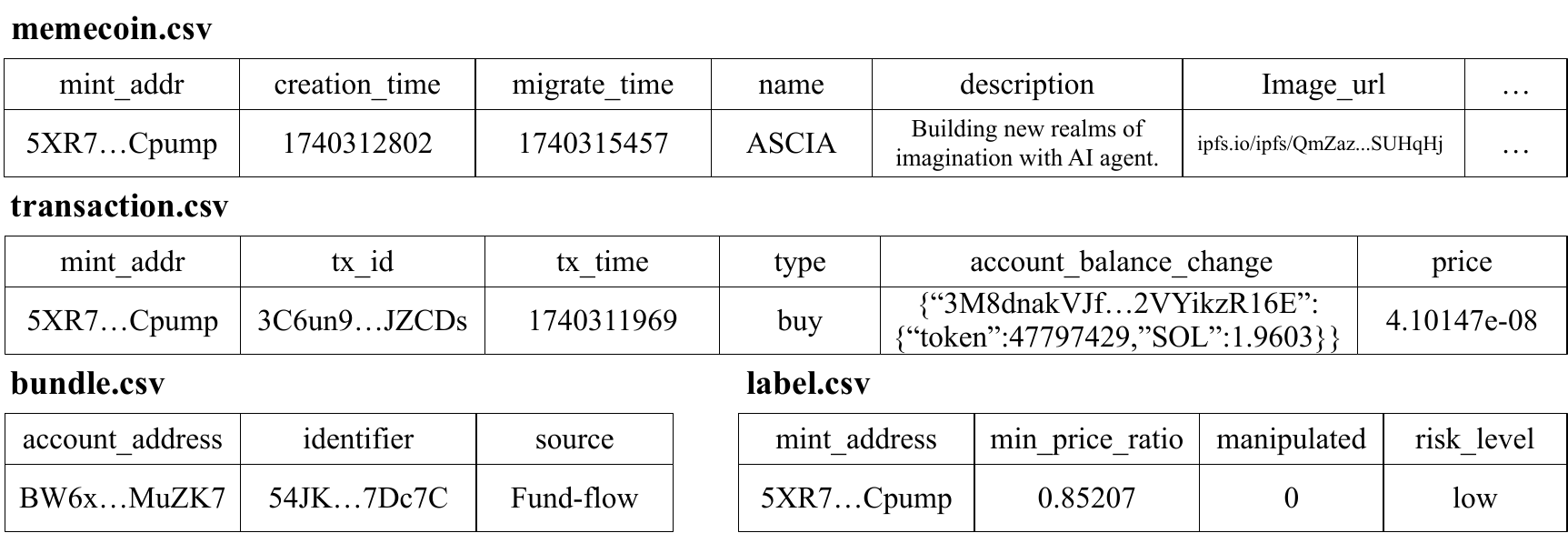}
\vspace{-0.8cm}
\label{tab:data_structure}
\end{table*} 

\vspace{-0.3cm}
\subsection{Memecoin Launch Collection}
\vspace{-0.2cm}

We identify memecoins that completed migration from Pump.fun to DEX by tracking transactions initiated by the \href{https://solscan.io/account/39azUYFWPz3VHgKCf3VChUwbpURdCHRxjWVowf5jUJjg}{Pump.fun creator account} that transfer SOL to the \href{https://solscan.io/account/7YttLkHDoNj9wyDur5pM1ejNaAvT9X4eqaYcHQqtj2G5}{\texttt{DEX fee account}}, which signals the creation of a new liquidity pool on DEX (migration). From the migration logs, we extract the mint address of the memecoin and the address of the DEX pool. For each mint address, we further retrieve token metadata (name, symbol, description, image, social handles) via the Metaplex Token Metadata program-derived address. The resulting memecoin launch records are summarized in memecoin.csv (Table~\ref{tab:data_structure}), with overall statistics reported in Table~\ref{tab:dataset_stats}.

\vspace{-0.3cm}

\subsection{Transaction-to-Behavior Extraction}
\vspace{-0.2cm}

For each memecoin mint, we collect on-chain transactions involving the memecoin. Pre-migration transactions span from creation to migration, and post-migration transactions cover the first hour after migration. Pre-migration data is used for feature engineering, while post-migration data is reserved for risk annotation.

Since raw transactions are heterogeneous at the function level, we introduce a parser that first identifies the interacting program (e.g., Pump.fun bonding curve, DEX liquidity pool, or aggregator routes) and counterparty addresses from transaction logs, then computes the net balance change of each involved account to determine token flow direction and magnitude. Based on these signals, we classify transaction type as \emph{create} (newly minted token), \emph{buy} or \emph{sell} (swap between a user and a liquidity pool), \emph{wash trade} (buy and sell within a single transaction), or \emph{transfer} (between user accounts). Table~\ref{tab:dataset_stats} reports the structural transaction records after extraction. Among 30.8M pre-migration transactions, 73.5\% are normal buy or sell, 21.4\% are wash trades, and 4.9\% are transfers. Notably, 98.7\% of "create" events co-occur with developer buys in the same transaction, indicating pervasive first-tier inventory pre-acquisition by developers. The post-migration 1-hour transaction count is approximately six times that of the entire pre-migration period, showing that migration to a DEX drastically amplifies market exposure.




\begin{table}[tbp]
\centering
\footnotesize
\caption{Statistics of collected memecoins and transactions.}
\begin{tabular}{lcccc}
\toprule
 & \textbf{Memecoins} & \textbf{Overall Tx.} & \textbf{Pre-mig. Tx.} & \textbf{Post-mig. Tx.} \\
\midrule
Total  \#           & 41,470 & 218,530,716 & 30,833,503 & 187,697,213 \\
Avg. \# per memecoin  & --     & 5,269       & 743        & 4,526       \\
Unique account \#    & --     & --          & 1,916,363  & 6,538,577   \\
\bottomrule
\end{tabular}
\vspace{-0.5cm}
\label{tab:dataset_stats}
\end{table}

\subsection{Bundle Trace Identification}


Insiders often hold tokens across multiple accounts to mask the true holding concentration. Identifying these bundled accounts allows us to reveal the actual holding concentration. We design heuristics to identify the bundle traces from three sources:

(1) \textbf{Multi-account co-purchase}: Some insiders use multiple addresses within the same transaction to purchase tokens directly. Since signing such a transaction requires the private keys of all participating accounts, we treat accounts that jointly execute buys or sells in the same transaction as belonging to the same entity.

(2) \textbf{Fund-flow relationship}: On Solana, creating a new account requires an initial deposit to cover rent, so accounts funded by the same funding address likely share common ownership. We identify the funding addresses of the 1,916,363 unique accounts involved in pre-migration transactions to reconstruct this relationship. We exclude funding addresses that belong to centralized exchanges, as their outgoing transfers typically reflect user withdrawals rather than coordinated control.

(3) \textbf{Jito bundle ID}: Jito is a widely used block relay on Solana that enables atomic bundled execution of transactions. Insiders frequently use Jito to bundle buy and sell transactions across multiple wallets, enabling fast, simultaneous multi-account actions. These relationships are not included in on-chain data. For 22,668,222 pre-migration transactions that are labeled as "buy" or "sell", we crawled their Jito bundle IDs from the \href{https://explorer.jito.wtf/}{Jito Explorer}. We treat accounts whose transactions share the same Jito bundle ID as controlled by the same entity.




\begin{table}[htbp]
\centering
\footnotesize
\vspace{-0.4cm}
\caption{Statistics of bundle data by source.}
\begin{tabular}{lcc}
\toprule
\textbf{Source} & \textbf{Bundled Holder Proportion} & \textbf{Bundled Supply Proportion} \\
\midrule
Co-purchase & 6.46\% & 9.16\%\\ 
Fund-flow & 22.57\%  & 28.22\% \\
Jito Bundle & 10.39\% & 15.96\% \\ 
\midrule
All& 28.13\% & 36.50\% \\
\bottomrule
\end{tabular}
\label{tab:bundle_statistics}
\vspace{-0.2cm}
\end{table}

As shown in Table~\ref{tab:data_structure}, a bundle trace is organized as a (user account, identifier) tuple, where the identifier can be a transaction ID, a funder address, or a Jito bundle ID. For each memecoin launch, we cluster user accounts into bundles by matching identifier values and merge overlapping bundles. Table~\ref{tab:bundle_statistics} reports the share of bundled accounts among holders and their share of total token supply. Notably, 36.5\% of the total token supply is held by bundled accounts at the point of migration, revealing pervasive multi-account concealment.

Bundle traces reveal the true holding concentration that raw on-chain data conceals. For example, after merging bundled accounts, the top-10 holder share increases by 24 percentage points for high-risk memecoins, compared to only 6 points for low-risk memecoins, indicating that bundle traces provide an informative signal for high-risk detection.



\subsection{Feature Engineering}
\label{sec:feature}
We construct 122 features across five categories computed 
strictly from pre-migration data to prevent label leakage:
\textbf{(1) contextual information} (e.g., SOL price, calendar time; 
6 features); 
\textbf{(2) holding concentration} (developer/fast-mover/top-holder 
shares; 59 features); 
\textbf{(3) market activity} (transaction counts, traders, volume; 
22 features); 
\textbf{(4) bundle statistics} (concentration after bundle-account 
clustering; 35 features); and 
\textbf{(5) time-series} (pre-migration price/volume dynamics; time\_span * 5 features). Table~\ref{tab:feature} summarizes these groups along with their representative features.

\begin{table*}[htbp]
\centering
\footnotesize
\caption{Summary of designed feature groups for memecoin launch.}
\label{tab:feature_summary}
\renewcommand{\arraystretch}{1.15}
\setlength{\tabcolsep}{5pt}

\rowcolors{3}{gray!6}{white}

\begin{tabular}{p{2.5cm}p{4.5cm}p{6cm}}
\toprule
\textbf{Feature Group} & \textbf{Description} & \textbf{Representative Features} \\ 
\midrule

\textbf{Contextual information} &
Launch-time context of the token (e.g., SOL price and calendar time). &
SOL\_price, migrate\_weekday, migrate\_hour, migrate\_month, \textit{etc.} \\ 
\textbf{Holding concentration} &
Holding concentration of tokens among early buyers (developer, first-movers). & 
dev\_hold\_pct, early\_top$k$\_pct, top$k$\_hold\_pct, \textit{etc.} \\ 
\textbf{Market activity} &
Pre-migration trading activity, including trader participation, volume, and wash trading signals. &
tx\_num, time\_span, trader\_num, holder\_num, buy\_num, sell\_num, \textit{etc.} \\ 

\textbf{Bundle statistics} &
Holding concentration and trading statistics after grouping bundled accounts. & bundle\_hold\_pct, bundle\_num, bundle\_account\_num, top$k$\_hold\_pct, early\_top$k$\_hold\_pct, \textit{etc.}  \\ 

\textbf{Time-series} &
Price and volume time-series between the creation and migration time. &
time\_span * [open\_price, end\_price, avg\_price, volume] \\ 

\bottomrule
\end{tabular}
\label{tab:feature}
\end{table*}

\subsection{Risk Level Annotation}
\label{sec:label}

We annotate each memecoin launch with a risk-level label using a two-layer approach that integrates a statistical label and a manual label, both derived from post-migration data.

\textbf{Statistical label}: Post-migration price movements reflect a tug-of-war between insider sell-offs and external buying pressure. However, a substantial price drop shortly after migration is most likely attributable to insiders unwinding their low-cost inventory rather than to organic market dynamics. Based on this intuition, we define \texttt{min\_price\_ratio} as the minimum token price observed within $y$ minutes after migration, normalized by the migration price. The choice of $y$ reflects a trade-off: too small a window risks missing insider sell-offs that are still in progress, while too large a window allows the natural decay of speculative attention to dominate the signal. Empirically, we find that $y = 20$ minutes provides a reasonable balance: as shown in Table~\ref{tab:min_ratio_bins}, about 73\% of memecoins drop below 40\% of their migration price within this window.

\begin{table}[t]
\centering
\footnotesize
\caption{Distribution of memecoins \textit{w.r.t.} \texttt{min\_price\_ratio}.}
\label{tab:min_ratio_bins}
\setlength{\tabcolsep}{5pt}
\begin{tabular}{lcccccc}
\toprule
\textbf{Value Range} & $[0.0, 0.2)$ & $[0.2, 0.4)$ & $[0.4, 0.6)$ & $[0.6, 0.8)$ & $[0.8, 1.0)$ & $[1.0, +\infty)$ \\
\midrule
\textbf{Memecoin \#}  & 24,988 & 5,265  & 3,315 & 2,991 & 2,756 & 2,155 \\
\textbf{Percentage}   & 60.26\% & 12.70\% & 7.99\% & 7.21\% & 6.65\% & 5.20\% \\
\bottomrule
\vspace{-0.6cm}
\end{tabular}
\end{table}

\begin{figure*}[tbp]
\centering
\includegraphics[width=14cm]{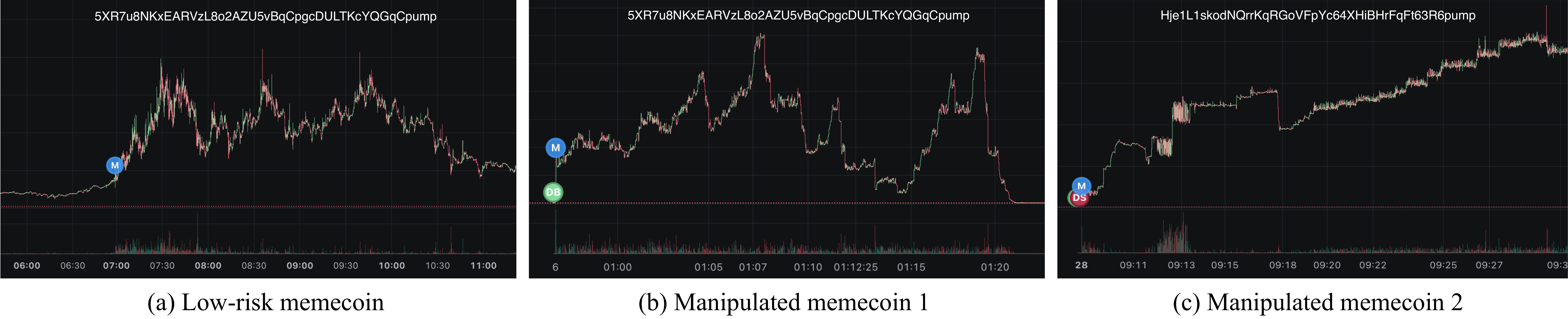}
\caption{Examples of a low-risk memecoin and two manipulated memecoins. For memecoin (b) and (c), the price move trajectories are artificial and controlled by manipulators with pre-programmed bots, and can therefore be readily identified by humans.} 
\vspace{-0.5cm}
\label{fig:manipulation_example}
\end{figure*}

\textbf{Manual label}: A short time window allows \texttt{min\_price\_ratio} to capture unambiguous insider sell-offs that occur shortly after migration, but is insufficient to capture cases where the dump unfolds over a longer horizon. A typical example is that insiders actively control prices to keep price artificially high through recurring sell-buy cycles, as illustrated in Figure~\ref{fig:manipulation_example}(b) and (c). To capture these cases that the statistical indicator alone cannot cover, we introduce a second layer of manual annotation: for each memecoin with \texttt{min\_price\_ratio} $\geq$ 0.5, we use the memecoin trading interface \href{https://gmgn.ai}{GMGN} to inspect each memecoin's full post-migration price-volume trajectory and screen whether it is manipulated. Among 9,471 such memecoins, 37.5\% are labeled as manipulated. 

This two-layer design balances annotation effort with coverage: the statistical indicator captures the bulk of clearly high-risk memecoins at scale, leaving only the ambiguous cases for human inspection. Finally, we introduce a default risk-level categorization, which can be easily configured depending on what level of risks that researchers aim to detect:

\begin{tcolorbox}[
    colback=gray!5,
    colframe=gray!50,
    boxrule=0.5pt,
    arc=2pt,
    left=8pt, right=8pt, top=6pt, bottom=6pt,
]
    
    \textbf{High-risk:} triggers any of \texttt{min\_price\_ratio} $< 0.3$ or \texttt{is\_manipulated}
    
    \textbf{Low-risk:} satisfies all of \texttt{min\_price\_ratio} $\geq 0.7$ and \texttt{not\_manipulated}
    
    \textbf{Medium-risk:} otherwise.
\end{tcolorbox}
Table~\ref{tab:label_annotation} reports the label distribution, from which we observe that the majority of tokens (84.13\%) fall into the high-risk region. 


\begin{table}[htbp]
\centering
\footnotesize
\vspace{-0.5cm}
\caption{Statistics of annotation.}
\label{tab:label_annotation}
\begin{tabular}{lccc}
\toprule
\textbf{Risk Level} & \textbf{High} & \textbf{Medium} & \textbf{Low} \\
\midrule
\textbf{Memecoin \#}      & 34,890 & 4,702 & 1,878 \\
\textbf{Percentage} & 84.13\% & 11.34\% & 4.53\% \\
\bottomrule
\end{tabular}
\vspace{-0.5cm}
\end{table}


\subsection{Findings}

\label{app:observation}
Based on the collected dataset, we observe that high-risk memecoins exhibit distinct patterns during the launchpad sale. 

\begin{figure}[tbp]
\centering
\includegraphics[width=14.3cm]{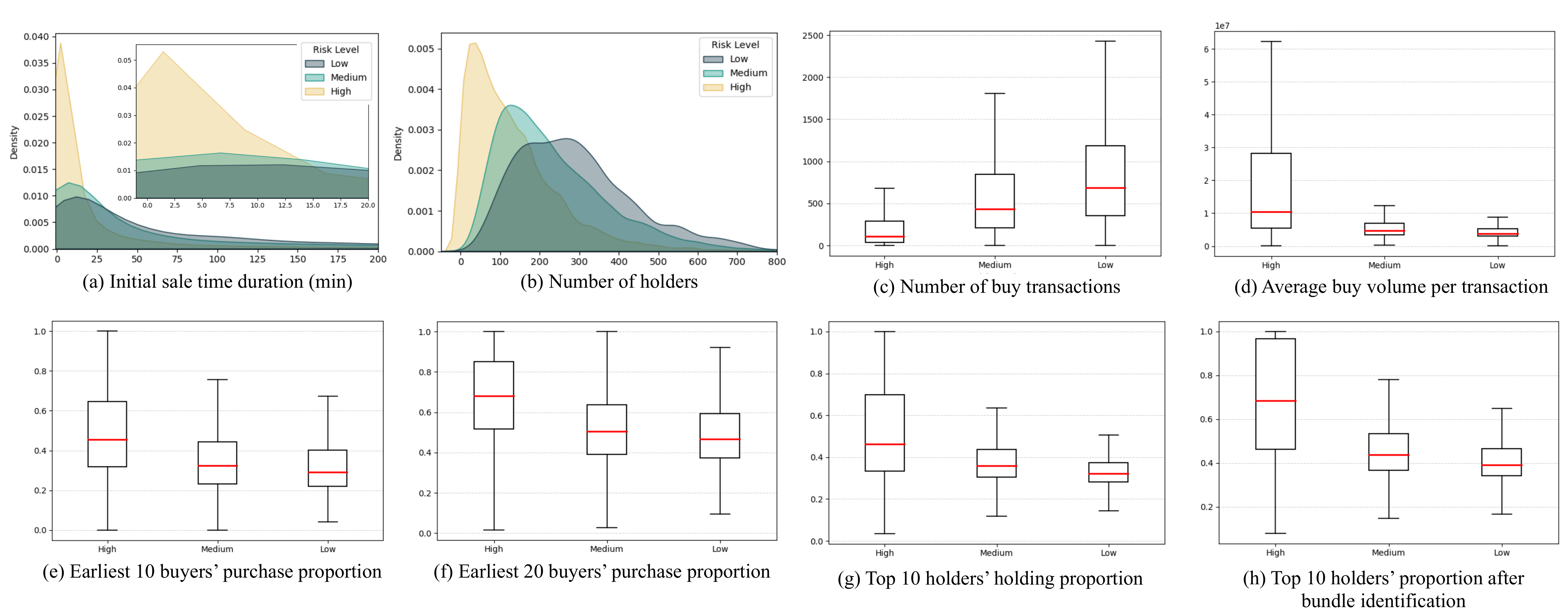}
\caption{High-risk memecoins exhibit shorter launchpad sale durations (a), fewer holders (b), and fewer buy transactions (c), while each buy involves larger token volumes (d). These statistics results from early buyers accumulating a large share of the token supply (e, f), highly concentrated holdings, which is more pronounced after bundle identification (g, h).} 

\label{fig:token_study}
\end{figure}

\textbf{Concentrated early accumulation:} First, as shown in Figure~\ref{fig:token_study}(e) and (f), the first buyers of high-risk tokens accumulated a much larger share of the supply than medium- and low-risk memecoins. According to statistics, the first 10 and 20 buyers of high-risk tokens hold 17 and 19 percentage points more of the supply than those of low-risk tokens. Consequently, when they reach the migration point, high-risk tokens tend to exhibit more concentrated holding distributions, as shown in Figure~\ref{fig:token_study}(g).

\textbf{Faster, thinner sales:} Second, since a substantial portion of the high-risk token supply has already been acquired by early buyers, it is easier and faster for high-risk tokens to reach the migration requirement. As a result, high-risk memecoin launches tend to have shorter time spans, as shown in Figure~\ref{fig:token_study}(a), and fewer buy transactions as shown in Figure~\ref{fig:token_study}. Meanwhile, the average buy volume of high-risk memecoins is much larger than that of low-risk memecoins, shown in Figure~\ref{fig:token_study}(d), and the number of holders at the migration point is smaller, as shown in Figure~\ref{fig:token_study}(b).

\textbf{Bundle-revealed concealment:} Finally, in Figure~\ref{fig:token_study}(h), we re-calculate the top-10 holding percentage after grouping shares into bundled accounts controlled by same users. Comparing Figure~\ref{fig:token_study}(g) and (h), we observe that the median holding percentages increase by 24, 9, and 6 percentage points for high-, medium-, and low-risk tokens, respectively. This suggests that insiders of high-risk tokens are more likely to use bundled accounts to conceal their true holding concentration.

\section{High-Risk Memecoin Detection}

We introduce a new detection task that predicts whether a newly migrated memecoin is high risk after migration. For this task, we label high-risk memecoins as category 1, and medium- and low-risk memecoins as category 0 for binary classification. In our experiments, we exclude memecoins with a launchpad sale duration shorter than one minute or fewer than 100 holders. The rule filters out 19,835 (47.8\%) memecoins with 95\% of them flagged as high-risk. This filtering mirrors simple screening rules commonly used by traders to discard very low-quality memecoins, and also helps mitigate class imbalance. Table~\ref{tab:task_statistics} summarizes the statistics of the dataset used in our task.

\textbf{Objective function}: We use a class-weighted binary cross-entropy loss:
\[
\mathcal{L}(\theta)
= - \sum_{i=1}^{N} 
\big[
w_1\, y_i \log p_\theta(x_i) 
+ 
w_0\, (1 - y_i) \log (1 - p_\theta(x_i))
\big],
\]
where $w_1$ and $w_0$ are class weights chosen to be inversely proportional to the number of samples in each class. This ensures that the minority class receives proportionally more weight during training, preventing the model from being dominated by the majority class.

\begin{table}[htbp]
\centering
\footnotesize
\vspace{-0.3cm}
\caption{Statistics of the subsampled dataset for experiments}
\begin{tabular}{lccc}
\toprule
\textbf{Risk Level} & \textbf{Memecoin \#} & \textbf{Percentage} & \textbf{Label} \\
\midrule
High & 16,048 & 74.18\% & 1 \\
Normal & 5,587 & 25.82\% & 0 \\
\bottomrule
\end{tabular}
\label{tab:task_statistics}
\end{table}

\subsection{Experimental Setup}
\label{sec:experiment_setup}

We split the dataset chronologically by each memecoin's creation time into training and testing sets at a 7:3 ratio. For all the approaches, we tune hyperparameters using five-fold cross-validation on the training set, and report the average test performance over five independent runs.

\textbf{Metrics:} We adopt precision, recall, and F1 scores for both classes and macro averaging to provide an overall view of model performance. Since accuracy is unreliable under class imbalance, we employ the Area Under the Precision-Recall Curve (AUPRC) to evaluate the model’s ability to distinguish high-risk from non-high-risk tokens under class imbalance. A higher AUPRC indicates that the model maintains high precision and recall simultaneously under skewed class distributions.

\textbf{Baselines:} We evaluate two model families: (i) tabular models on Groups 1–4, including Logistic Regression (LR), Random Forest (RF)~\citep{breiman2001random} , LightGBM~\citep{lightgbm}, XGBoost~\citep{chen2016xgboost}, and Multi-Layer Perceptron (MLP); and (ii) time-series model on Group 5 (time-series), including TCN~\citep{tcn}, LSTM~\citep{lstm}, GRU~\citep{gru} and Transformer encoder~\citep{transformer}. Hyperparameters are in Appendix~\ref{app:hyper_parameter}.

We do not adopt graph models because our task is memecoin-level detection, not account-level tasks such as Sybil detection. Although bundle traces are relational at the account level (forming fully connected subgraphs), we have distilled them into memecoin-level statistical features (feature group 4) that reveal the true holding concentration. The ablation in Section~\ref{sec:feature_ablation} confirms that this distillation captures the predictive signal in bundle structure.

\begin{table*}[tbp]
\footnotesize
\centering
\caption{Performance comparison on high-risk memecoin detection.}
\begin{tabular}{@{}l@{\hspace{10pt}}cccccccccc@{}}
\toprule
\multirow{2}{*}{\textbf{Model}} & 
\multirow{2}{*}{\textbf{AUPRC}} & 
\multicolumn{3}{c}{\textbf{Label=0} (normal)} & 
\multicolumn{3}{c}{\textbf{Label=1} (high-risk)} & 
\multicolumn{3}{c}{\textbf{Macro}} \\
\cmidrule(lr){3-5} \cmidrule(lr){6-8} \cmidrule(lr){9-11}
& & Precision & Recall & F1 & Precision & Recall & F1 & Precision & Recall & F1  \\
\midrule
Random  & 0.2589 & 0.2589 & 0.2589 & 0.2589 & 0.7411 &  0.7411 &  0.7411 & 0.5000 & 0.5000 & 0.5000 \\
\midrule
LR    & 0.5338 & 0.5122 & 0.5441 & 0.5277 & 0.8358 & 0.8175 & 0.8265 & 0.6740 & 0.6808 & 0.6771 \\
RF    & 0.5688 & 0.5444 & \textbf{0.5766} & \textbf{0.5600} & 0.8477 & 0.8300 & 0.8387 & \textbf{0.6960} & \textbf{0.7033} & \textbf{0.6994} \\
XGBoost &  0.5636 & 0.5264 & 0.5482 & 0.5371 & 0.8385 & 0.8263 & 0.8323 & 0.6824 & 0.6872 & 0.6847 \\
LGBM  & 0.5642 & 0.5360 & 0.5452 & 0.5406 & 0.8388 & 0.8337 & 0.8363 & 0.6874 & 0.6895 & 0.6884 \\
MLP & \textbf{0.5729} & \textbf{0.5450} & 0.5695 & 0.5570 & 0.8459 & \textbf{0.8325} & \textbf{0.8391} & 0.6954 & 0.7010 & 0.6981 \\
\midrule
GRU & 0.4972 & 0.4913 & 0.5186 & 0.5046 & 0.8270 & 0.8108 & 0.8189 & 0.6567 & 0.6680 & 0.6613\\
LSTM & 0.5023 & 0.4895 & 0.5393 & 0.5132 & 0.8317 & 0.8019 & 0.8165 & 0.6606 & 0.6706 & 0.6649 \\
TCN & 0.4844 & 0.4847 & 0.5074 & 0.4958 & 0.8236 & 0.8100 & 0.8167 & 0.6542 & 0.6587 & 0.6563 \\
Transformer & 0.4841 & 0.4658 & 0.4920 & 0.4786 & 0.8174 & 0.8013 & 0.8093 & 0.6416 & 0.6466 & 0.6439 \\
\midrule
MLP+RF & 0.5804 & 0.5572 & 0.5701 & 0.5636 & 0.8473 & 0.8404 & 0.8438 & 0.7023 & 0.7052 & 0.7037\\
MLP+LGBM & 0.5821 & 0.5561 & 0.5683 & 0.5622 & 0.8467 & 0.8402 & 0.8435 & 0.7014 & 0.7043 & 0.7028 \\
MLP+LSTM & 0.5827 & 0.5490 & 0.5730 & 0.5607 & 0.8466 & 0.8427 & 0.8446 &  0.6958 & 0.7079 & 0.7027 \\
\bottomrule
\label{tab:comparison}
\vspace{-0.4cm}
\end{tabular}
\end{table*}

\begin{table*}[tbp]
\centering
\footnotesize
\caption{Feature ablation study. Group 1, 2, 3, 4 corresponds to contextual information (Ctx), holding concentration (Hold), market activity (Mkt), and bundle statistics (Bnd).}
\label{tab:ablation_study}
\begin{tabular}{lccccccccccc}
\toprule
\multirow{2}{*}{\textbf{Feature}} & 
\multirow{2}{*}{\textbf{AUPRC}} & 
\multicolumn{3}{c}{\textbf{Label=0}} & 
\multicolumn{3}{c}{\textbf{Label=1}} & 
\multicolumn{3}{c}{\textbf{Macro}} \\
\cmidrule(lr){3-5} \cmidrule(lr){6-8} \cmidrule(lr){9-11}
& & Prec. & Rec. & F1 & Prec. & Rec. & F1 & Prec. & Rec. & F1  \\
\midrule
Full  & 0.5729 & 0.5450 & 0.5695 & 0.5570 & 0.8459 & 0.8325 & 0.8391 & 0.6954 & 0.7010 & 0.6981 \\
\midrule
– Ctx & 0.5567 & 0.5216 & 0.5565 & 0.5385 & 0.8400 & 0.8202 & 0.8300 & 0.6808 & 0.6883 & 0.6842 \\
– Hold & 0.5693 & 0.5358 & 0.5582 & 0.5468 & 0.8420 & 0.8296 & 0.8358 & 0.6889 & 0.6939 & 0.6913 \\
– Mkt  & 0.5369 & 0.5165 & 0.5452 & 0.5305 & 0.8366 & 0.8202 & 0.8283 & 0.6766 & 0.6827 & 0.6794 \\
– Bnd  & 0.5451 & 0.5251 & 0.5492 & 0.5369 & 0.8349 & 0.8287 & 0.8318 & 0.6800 & 0.6890 & 0.6844 \\
– Hold\&Bnd & 0.5268 & 0.5048 & 0.5257 & 0.5151 & 0.8304 & 0.8183 &  0.8243 & 0.6676 & 0.6720 & 0.6697 \\
\bottomrule
\end{tabular}
\vspace{-0.4cm}
\end{table*}

\subsection{Performance Comparison}

We evaluate existing approaches and report their performance in Table~\ref{tab:comparison}, where we make several observations: (i) Tabular feature-based models achieve better performance compared with time-series models, suggesting that designed features are more informative for detecting high-risk memecoins than pure time-series of price and trading volumes. (ii) Among the tabular models, MLP achieves the best performance, likely due to its ability to implicitly combine multiple features and capture higher-order interactions beyond the original feature set. (iii) Among time-series models, more complex architectures such as Transformers do not achieve the best performance, whereas LSTM performs the best. We conjecture that this is because the short and noisy trading sequences lack the long-range dependencies that Transformers are designed to exploit.

\subsection{Feature Ablation Study}
\label{sec:feature_ablation}
To measure the contribution of features, we iteratively remove each feature group and report the results in Table~\ref{tab:ablation_study}. We first observe that removing any feature group results in a performance drop, indicating that the designed features provide complementary information. Second, removing Group 3 (market activity) leads to the largest performance degradation. This observation is consistent with the feature importance scores obtained from the RF model (see Figure~\ref{fig:importance} in Appendix~\ref{app:feature_importance}), where features in the market activity group receive the highest importance values. Moreover, groups 2 and 4 both capture holding concentration, but Group 4 is computed after leveraging bundle traces to reveal true ownership concentration. Consequently, removing Group 4 results in a larger performance drop than removing Group 2. Since these groups capture complementary aspects of holding concentration, removing both leads to a substantial performance degradation.

\subsection{Application: Memecoin Selection}

To assess whether risk detection models can reduce financial loss in practice, we simulate a simple memecoin selection strategy by ranking tokens based on predicted risk scores in descending order and selecting the top-$k$ candidates. For each selected memecoin, we use the price at migration as the purchase price and compute the average loss by randomly sampling 100 selling timestamps within the following hour. We evaluate this strategy on the test set and report the results in Table~\ref{tab:token_selection}. Precision measures the proportion of non-high-risk memecoins among the selected tokens, while loss denotes the percentage loss incurred when investing one unit of capital. As shown, random selection leads to losses exceeding 60\%, whereas model-guided selection reduces losses to around 30\%, despite the models not being optimized for loss minimization. Among all approaches, MLP achieves the best performance, reducing financial loss by 34 percentage points. These results demonstrate that MELT provides practical value for real-world loss mitigation.

\begin{table}[tbp]
\centering
\footnotesize
\caption{Performance of memecoin selection.}
\begin{tabular}{lcccc}
\toprule
\multirow{2}{*}{\textbf{Model}} 
& \multicolumn{2}{c}{\textbf{Top-100}} 
& \multicolumn{2}{c}{\textbf{Top-200}} \\
\cmidrule(lr){2-3} \cmidrule(lr){4-5}
& \textbf{Precision} 
& \textbf{Loss (\%)} 
& \textbf{Precision} 
& \textbf{Loss (\%)} \\
\midrule
w.o. model & 0.2540 & 60.71 & 0.2650 & 64.84 \\                 
\midrule
LR & 0.7000 & 34.06 & 0.6200 & 36.50 \\
RF & 0.7600 & 30.34  & 0.7300 & 33.72   \\
XGBoost  & 0.7500 & 31.45  & 0.7400 & 33.15 \\
LGBM  & 0.7400 & 31.99 & 0.7250 & 34.15 \\
MLP  & 0.8000 & 26.64  & 0.7550 & 31.47  \\
\bottomrule
\label{tab:token_selection}
\vspace{-0.7cm}
\end{tabular}
\end{table}

\vspace{-0.2cm}
\section{Related Work}
\vspace{-0.2cm}
\textbf{Rug Pull Schemes.} Rug pull refers to an exit scam pattern on blockchain, where token creators initially provide liquidity to a token pair on a DEX, attract investor trading activity, and later abruptly remove the liquidity, draining the pool of valuable assets and rendering the token worthless. The attack is achieved via the creator’s full control over liquidity pool tokens. Rug pulls have been the subject of extensive analysis in the blockchain security community. Cernera et al.~\citep{cernera2023token} conduct a longitudinal analysis of over 1.3 million tokens on Ethereum and BNB Smart Chain, introducing the notion of 1-day rug pulls, where a token is listed and drained within 24 hours. Their methodology relies on liquidity pool event patterns and identifies sniper bots that assist in early exploitation. Mazorra et al.~\citep{mazorra2022not} build a supervised classifier using blockchain and social features, showing that most rug pulls are initiated by wallets with prior history of creating disposable tokens. Yaremus et al.~\citep{yaremus2025detecting} construct a taxonomy of exit scams and propose an automated labeling pipeline based on on-chain liquidity events. Other datasets like SolRPDS~\citep{alhaidari2024solrpds} provide curated rug pull samples on Solana to facilitate model training.

\textbf{Pump-and-dump Schemes.} Pump-and-dump campaigns coordinate buying to create a short-lived price spike that enables early insiders to sell at inflated prices~\citep{hu2023sequence,li2021cryptocurrency,xu2019anatomy,la2021doge}. Prior work has mainly studied such behavior in centralized exchanges (CEXs) or order-book markets, where transaction-level traces are not publicly accessible and holdings are custodial rather than address-based, making it difficult to reconstruct fine-grained trading traces or analyze multi-account behaviors~\citep{kamps2018moon,hu2023sequence}. Moreover, order-book price dynamics are driven by buy orders lifting the ask side and subsequent sell orders consuming liquidity. In contrast, launchpad-based memecoin markets sell tokens via a bonding-curve program and then migrate liquidity to a DEX pool, which confers a much stronger first-mover advantage to early buyers and can amplify post-migration losses when these low-cost positions unwind.

\vspace{-0.2cm}
\section{Discussion}
\vspace{-0.2cm}
\label{sec:discussion}
\textbf{Limitations.}  Our study focuses on memecoins that successfully migrated to a DEX. While migrated memecoins attract the vast majority of trading attention, unmigrated memecoins can still draw retail buyers during the launchpad sale and cause financial losses at smaller scales. Extending MELT to cover the long tail of unmigrated launches is left as future work.

\textbf{Utility Outlook.} Beyond studying high-risk memecoins, MELT offers broader utility in the crypto ecosystem. For example, the bundle trace data enables research on linking accounts belonging to the same entities, revealing hidden coordination behaviors on-chain such as Sybil or money laundering detection~\citep{hu2025matching,bert4eth,beres2021blockchain}. Both the dataset and the released data science pipeline can be extended to continuously collect newly launched memecoins, perform large-scale transaction collection and parsing, and identify bundled or coordinated accounts. The rapid rise of the Solana ecosystem introduces new research challenges in large-scale memecoin markets, making MELT a timely resource for future research and the ML community.

\vspace{-0.3cm}
\section{Conclusion}
\vspace{-0.2cm}

The rise of launchpads has transformed how memecoins are issued, and also created a new class of threat whose signals lie deep within early trading behaviors. To open this problem to systematic study, we presented MELT, a behavioral trace dataset that uncovers and reconstructs launchpad-stage behavioral traces from heterogeneous on-chain and off-chain signals, and connects them to the post-migration risk level exhibited by a memecoin once it enters the public DEX. Experiments show that incorporating prediction signals learned from MELT substantially reduces investment loss, confirming it carries practical defensive value. We hope MELT contributes to the ML community as a resource for studying complex, adversarial user behaviors in real-world on-chain markets.

\bibliographystyle{plainnat}
\bibliography{main}

{
\small

\appendix

\section{Technical appendices and supplementary material}

\subsection{Feature Importance Study}
\label{app:feature_importance}

In Figure~\ref{fig:importance}, we report the importance scores of the 122 features produced by the RF model. As shown, features from Group~3 (market activity) and Group~4 (bundle statistics), highlighted in green and red, account for the highest importance, which is consistent with the feature ablation results in Table~\ref{tab:ablation_study}.

\begin{figure*}[htbp]
\centering
\includegraphics[width=14cm]{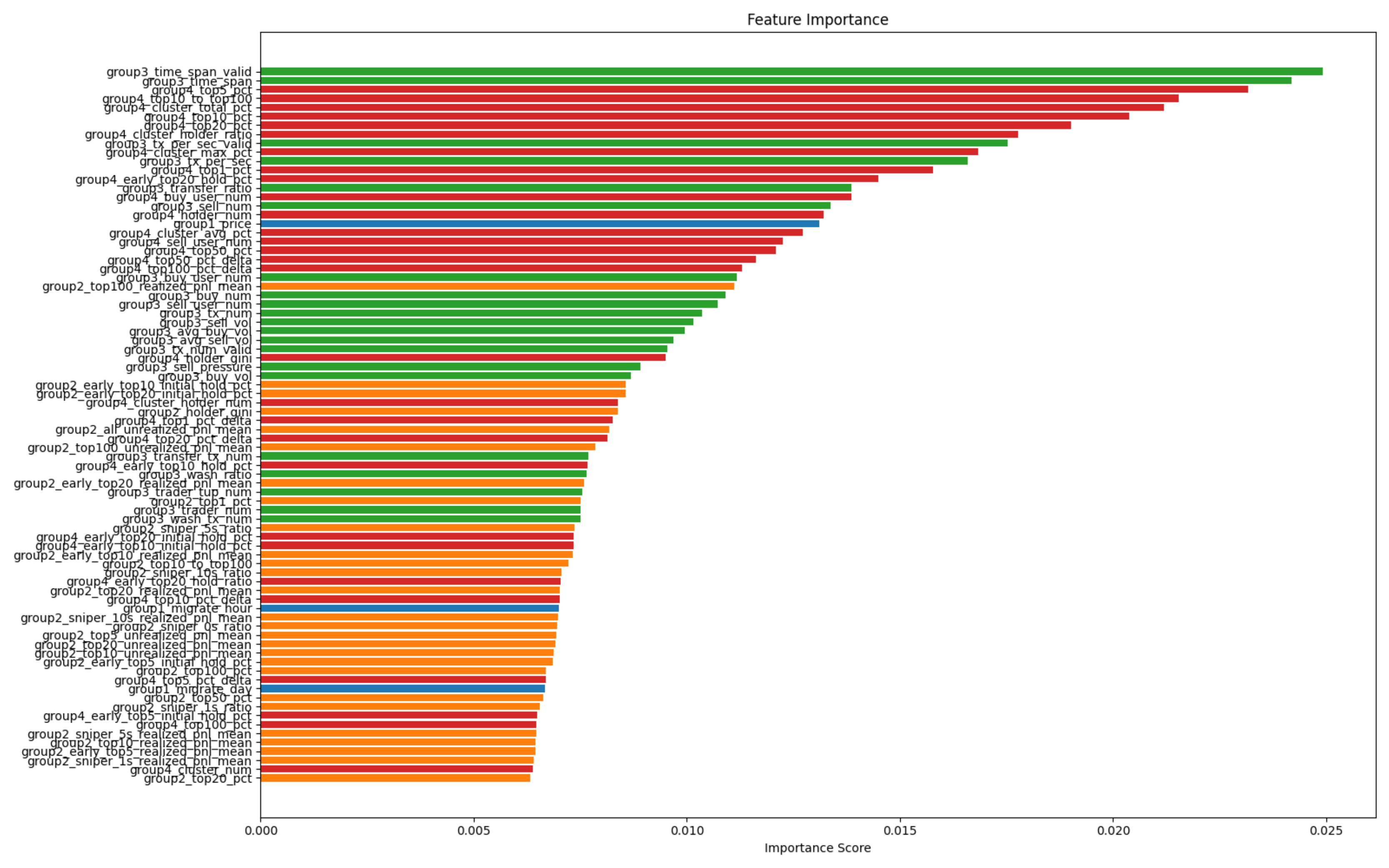}
\caption{Feature importance score calculated by the RF model. Blue marks the contextual information features, orange marks holding concentration features, green marks market activity features, and red marks the bundle statistic features.}
\label{fig:importance}
\end{figure*}

\subsection{Hyper-parameter Settings}
\label{app:hyper_parameter}

For the high-risk detection task, competitors are configured as follows: LR serves as a linear baseline with L2 regularization and a fixed regularization strength of 1.0. For tree-based models, RF is implemented with 800 trees and a maximum depth of 16, while other parameters follow standard practice. XGBoost uses 800 trees with a maximum depth of 6 and a learning rate of 0.05. LightGBM is trained with 2000 boosting iterations and a learning rate of 0.02, with the maximum number of leaves set to 64 and subsampling applied to control model complexity. MLP consists of two fully connected hidden layers with 512 and 256 units, respectively. For time-series models, both GRU, LSTM and Transformer classifiers are configured with two stacked layers and a hidden dimension of 256. The TCN consists of 8 dilated convolutional blocks (kernel size 9, dilations [1, 2, 4, \ldots, 128]. For all neural networks, each layer is followed by batch normalization, ReLU activation, and dropout with a rate of 0.2.

\end{document}